\documentclass[12pt,a4paper,titlepage]{article}
\usepackage{amsmath,color}
\usepackage{a4wide}
\usepackage[utf8]{inputenc}
\usepackage{booktabs}
\usepackage[title]{appendix}
\usepackage{mathrsfs}
\usepackage{float}
\usepackage{multirow}
\usepackage{graphicx,caption,subcaption}
\usepackage[space]{grffile}

\renewcommand{\eqref}[1]{(\ref{#1})}

\definecolor{hyprf}{cmyk}{1,0.5,0,0}

\usepackage[utf8]{inputenc}
\usepackage[top=50pt,bottom=60pt,left=68pt,right=66pt]{geometry}
\usepackage{amsmath}
\usepackage{booktabs}
\usepackage{amssymb}
\usepackage[title]{appendix}
\usepackage{mathrsfs}
\usepackage{float}
\usepackage{xcolor}
\usepackage{multirow}
\usepackage{graphicx,caption,subcaption}
\usepackage{comment}
\usepackage[space]{grffile}

\def\un#1{\relax\ifmmode\@@underline#1\else
        $\@@underline{\hbox{#1}}$\relax\fi}

\let\du=\du                     

\def\c{\chi}

\def\e{\epsilon}

\def\m{\mu}
\def\n{\nu}

\def\s{\sigma}

\def\z{\zeta}
\def\D{\Delta}

\def\G{\Gamma}

\def\O{\Omega}

\def\bo{{\raise-.3ex\hbox{\large$\Box$}}}               
\def\pa{\partial}                                       
\def\TH{{\raise.2ex\hbox{$\displaystyle \bigodot$}\mskip-4.7mu \llap H \;}}
\def\face{{\raise.2ex\hbox{$\displaystyle \bigodot$}\mskip-2.2mu \llap {$\ddot
        \smile$}}}                                      

\def\abs#1{\left| #1\right|}                    
\def\leftrightarrowfill{$\mathsurround=0pt \mathord\leftarrow \mkern-6mu
        \cleaders\hbox{$\mkern-2mu \mathord- \mkern-2mu$}\hfill
        \mkern-6mu \mathord\rightarrow$}
\def\dvec#1{\vbox{\ialign{##\crcr
        \leftrightarrowfill\crcr\noalign{\kern-1pt\nointerlineskip}
        $\hfil\displaystyle{#1}\hfil$\crcr}}}           

\def\frac#1#2{{\textstyle{#1\over\vphantom2\smash{\raise.20ex
        \hbox{$\scriptstyle{#2}$}}}}}                   
\def\sfrac#1#2{{\vphantom1\smash{\lower.5ex\hbox{\small$#1$}}\over
        \vphantom1\smash{\raise.4ex\hbox{\small$#2$}}}} 
\def\bfrac#1#2{{\vphantom1\smash{\lower.5ex\hbox{$#1$}}\over
        \vphantom1\smash{\raise.3ex\hbox{$#2$}}}}       
\def\afrac#1#2{{\vphantom1\smash{\lower.5ex\hbox{$#1$}}\over#2}}    

\def\[{\lfloor{\hskip 0.35pt}\!\!\!\lceil}
\def\]{\rfloor{\hskip 0.35pt}\!\!\!\rceil}

\def\du#1#2{_{#1}{}^{#2}}

\def\un{\underline}
\def\fracmm#1#2{{{#1}\over{#2}}}

\def\low#1{{\raise -3pt\hbox{${\hskip 0.75pt}\!_{#1}$}}}

\newskip\humongous \humongous=0pt plus 1000pt minus 1000pt

\newif\ifdtup

\def\({\left(}
\def\){\right)}

\def\beq{\begin{equation}}
\def\eeq{\end{equation}}
\def\bea{\begin{eqnarray}}
\def\eea{\end{eqnarray}}

\newcommand{\be}{\begin{equation}}
\newcommand{\ee}{\end{equation}}
\newcommand{\nbe}{\begin{equation*}}
\newcommand{\nee}{\end{equation*}}

\newcommand{\lb}{\label}

\begin{document}
\renewcommand{\arraystretch}{1.3}

\thispagestyle{empty}

\noindent {\hbox to\hsize{
\vbox{\noindent July 2024 \hfill IPMU24-0001 }}
\noindent  $~$  version 3\hfill }

\noindent
\vskip2.0cm
\begin{center}

{\Large\bf Improved model of  large-field inflation with primordial \\ 
\vglue.1in black hole production in Starobinsky-like supergravity}

\vglue.3in

Ryotaro Ishikawa~${}^{a,\&}$ and Sergei V. Ketov~${}^{a,b,c,\#,}$\footnote{The corresponding author} 
\vglue.3in

${}^a$~Department of Physics, Tokyo Metropolitan University\\
1-1 Minami-ohsawa, Hachioji-shi, Tokyo 192-0397, Japan \\
${}^b$~Research School of High-Energy Physics, Tomsk Polytechnic University\\
Tomsk 634028, Russian Federation\\
${}^c$~Kavli Institute for the Physics and Mathematics of the Universe (WPI)
\\The University of Tokyo Institutes for Advanced Study, Kashiwa 277-8583, Japan\\
\vglue.1in

${}^{\&}$~ishikawa-ryotaro@ed.tmu.ac.jp, ${}^{\#}$~ketov@tmu.ac.jp
\end{center}

\vglue.3in

\begin{center}
{\Large\bf Abstract}
\end{center}
\vglue.2in

A viable model of large-field (chaotic) inflation with efficient production of primordial black holes is proposed in Starobinsky-like (modified) supergravity leading to the "no-scale-type" K\"ahler potential and the Wess-Zumino-type ("renormalizable") superpotential. The cosmological tilts are in good (within $1\sigma$) agreement with Planck measurements of the cosmic microwave background radiation. In addition, the power spectrum of scalar perturbations has a large peak at smaller scales, which leads to
a production of primordial black holes from gravitational collapse of large perturbations with the masses about $10^{17}$ g. The
masses are beyond the Hawking (black hole) evaporation limit of $10^{15}$ g, so that those primordial black holes may be viewed as viable candidates for a significant part or the whole of the current dark matter. The parameters of the superpotential were fine-tuned for those purposes, while the cubic term in the superpotential is essential whereas the quadratic term should vanish. The vacuum after inflation (relevant to reheating) is Minkowskian. The energy density fraction of the gravitational waves induced by the production of
primordial black holes and their frequency were also calculated in the second order with respect to perturbations.

\newpage

\section{Introduction}

Starobinsky model of inflation \cite{Starobinsky:1980te} based on modified gravity with the higher spacetime scalar curvature against the Einstein-Hilbert action remains popular today despite its origin in 1980. It is because of its conceptual attractiveness (using only gravitational interactions consistent with General Relativity), excellent agreement with recent measurements of Cosmic Microwave Background (CMB) radiation  \cite{BICEP:2021xfz,Tristram:2021tvh},  and its simplicity (having only one free parameter that becomes fixed by the known amplitude of CMB scalar perturbations).

The Starobinsky model is distinguished in modified gravity because the $R^2$-gravity action (with a positive dimensionless coefficient in four spacetime dimensions) dominating over the Einstein-Hilbert action during inflation is the only ghost-free action in a quadratically generated gravity, being also  scale-invariant, which is the origin of flatness of the effective inflaton scalar potential during slow-roll inflation
\cite{Ketov:2021fww}. The inflaton scalar in the Starobinsky model is a physical mode of the higher-derivative gravity, which becomes manifest after a transformation of the Starobinsky gravity to the Einstein frame (scalar-tensor gravity). Hence,  Starobinsky's inflaton can be viewed as the Nambu-Goldstone mode associated with spontaneous breaking of scale-invariance of the $R^2$ gravity during inflation \cite{Ketov:2012jt}. It implies the $R^2$-term with a proper coefficient must be present in any viable  model of inflation based on modified $F(R)$ gravity. The magnitude of the inflationary scale predicted by the Starobinsky model is 
approximately five orders lower than the Planck scale that is the ultra-violet cutoff scale in the Starobinsky model, which assures stability of inflation against quantum gravity corrections, see the reviews \cite{Ketov:2012yz,Ketov:2019toi,Ketov:2021fww} for details.

The Hubble value $H \sim {\cal O}(10^{14})$ GeV during Starobinsky inflation is far beyond the electro-weak scale. On the  Hubble scale gravity becomes truly important and (presumably) unified with other fundamental physical interactions based on supergravity  \cite{Ellis:2013xoa}. Viable extensions of the Starobinsky model of inflation in modified gravity to modified supergravity were pioneered in Refs.~\cite{Gates:2009hu,Ketov:2010qz,Farakos:2013cqa,Cecotti:2014ipa}. Those supergravity models can be upgraded in order to include a  formation of Primordial Black Holes (PBH) during Starobinsky inflation, while those PBH may be an essential part (or the whole) of the current dark matter (DM), see Refs.~\cite{Carr:2020gox,Ketov:2023ykf} for a review. In particular, the Cecotti-Kallosh supergravity model \cite{Cecotti:2014ipa} of inflation in the Starobinsky-like supergravity was generalized in Ref.~\cite{Aoki:2022bvj} by adding a holomorphic function to the superpotential, which led to an efficient production of PBH with the masses beyond the Hawking (black hole) evaporation limit of $10^{15}$ g, though in marginal agreement with the 
Planck-measured CMB tilt of scalar perturbations within $3\sigma$ but outside $1\sigma$ accuracy. On the other hand, when demanding good (within $1\s$) agreement with CMB, the masses of produced PBH appeared to be under the Hawking limit in the supergravity model of Ref.~\cite{Aoki:2022bvj} and, hence, those PBH would not survive at present. In this paper we found a remedy to this problem.

The paper is organized as follows. The relevant background information available in the literature is summarized in Section 2 that represents our setup. Our new model is introduced in Section 3. Our results for inflation and primordial black holes production are given in Section 4. The gravitational waves induced by the PBH production are studied in Section 5. Our conclusion is Section 6.

Our paper is based on a different ansatz for the superpotential when compared to that of Ref.~\cite{Aoki:2022bvj}, and our new results are not an extension or an upgrade of the previously obtained ones, though we have used the same methods. We skipped a historical overview and many standard equations describing inflation and PBH together with observational constraints on them because of numerous comprehensive reviews available in the literature, and focused on our new results with a limited list of references. We use the units with  $\hbar=c=M_{\rm Pl}=1$ throughout this paper, where $M_{\rm Pl}=(8\pi G_{\rm N})^{-1/2}$ denotes the reduced Planck mass.

\section{Setup and background state-of-the-art}

Modified gravity theories are generally-covariant non-perturbative extensions of Einstein-Hilbert (EH) gravity theory by the higher-order terms. The higher order terms are irrelevant in the Solar system but are relevant in the high-curvature regimes (inflation, black holes) and for large cosmological distances (dark energy). Any modified gravity Lagrangian has the higher-derivatives and generically suffers from Ostrogradsky instability and ghosts. However, there are exceptions. In a quadratically generated gravity with respect to the spacetime curvature, the only ghost-free term is given by the Ricci scalar squared, $R^2$, with a positive coefficient. It leads to the Starobinsky model of modified gravity with the action
\be  \lb{stara}
S_{\rm Star.} = \fracmm{1}{2}\int \mathrm{d}^4x\sqrt{-g} \left( R +\fracmm{1}{6M^2}R^2\right)=\int \mathrm{d}^4x\sqrt{-g} ~F(R)~,
\ee
having the only parameter $M\approx 1.3\cdot 10^{-5}$ whose value is fixed by the CMB amplitude (WMAP normalization).~\footnote{We use the spacetime signature $(-,+,+,+,)$.} 

In the high-curvature regime relevant to inflation, the EH term can be ignored and the pure $R^2$-action becomes scale-invariant. The Starobinsky gravity has the special attractor solution in the Friedmann-Lemaitre-Robinson-Walker (FLRW) universe with the Hubble function \cite{Starobinsky:1980te}
\be 
 H(t) \approx \left (\fracmm{M}{6}\right)^2 (t_{\rm end}-t)~,
\ee
for $M (t_{\rm end}-t)\gg 0$. This solution spontaneously breaks the scale invariance of the $R^2$-gravity and, hence, implies
the existence of the associated Nambu-Goldstone boson called scalaron. Scalaron can be revealed by rewriting the Starobinsky
action to the quintessence form by the field redefinition (see, e.g., Ref.~\cite{Maeda:1988ab})
\begin{equation}  \lb{trans}
  \varphi =  \sqrt{ \fracmm{3}{2}} \ln F'(\c)   \quad {\rm and}\quad g_{\m\n}\to 2F'(\chi) g_{\m\n}~,
  \quad \chi=R~,
  \ee
which leads to
\be \lb{quint}
S[g_{\m\n},\varphi]  = \fracmm{1}{2}\int \mathrm{d}^4x\sqrt{-g} R 
 - \int \mathrm{d}^4x \sqrt{-g} \left[ \frac{1}{2}g^{\m\n}\pa_{\m}\varphi\pa_{\n}\varphi
 + V(\varphi)\right]~,\nonumber
\ee
with the potential 
\be \lb{starp}
V(\varphi) = \fracmm{3}{4}M^2\left[ 1- \exp\left(-\sqrt{\frac{2}{3}}\varphi\right)\right]^2~.
\ee
This potential is suitable for describing slow-roll inflation with scalaron $\varphi$ as the canonical inflaton of mass $M$ due to the infinite plateau with a positive height, and leads to the well-known predictions for the cosmological tilt $n_s$ of scalar perturbations and the tensor-to-scalar ratio $r$ for CMB,
\be \lb{tilts}
n_s\approx 1 - \fracmm{2}{N}~,\quad r \approx \fracmm{12}{N^2}~~,
\ee
which fit current CMB measurements \cite{BICEP:2021xfz,Tristram:2021tvh}
\be \lb{cmbt}
n_s = 0.9649 \pm 0.0042~(68\%~{\rm CL}) \quad {\rm and} \quad r< 0.032~(95\%~{\rm CL})~,
\ee
for the e-foldings number $N=55\pm 5$, defined by $N(t)=\int^{t_{\rm end}}_{t} H(\tilde{t}) d\tilde{t}$.

A locally supersymmetric extension of the action (\ref{stara}) is called Starobinsky (modified) supergravity. It has only the first and second powers of 
$R$ but also includes other fields of the supergravity multiplet, see Ref.~\cite{Ketov:2013dfa} for its general action. Similarly to Eq.~(\ref{trans}), the
Starobinsky supergravity action can be transformed to the (dual or equivalent) standard Einstein supergravity action that has only the first power of $R$ and is coupled to two chiral matter supermultiplets \cite{Cecotti:1987sa}. Those chiral matter supermultiplets are called $T$ that includes a complex physical scalar field (real inflaton and its pseudo-scalar superpartner called sinflaton), and $C$ that includes goldstino arising due to spontaneous supersymmetry breaking during inflation. Goldstino is eaten up by gravitino that becomes massive (super-Higgs effect). The K\"ahler potential and the superpotential of the matter $(T,C)$ have the special ("no-scale") structure that can be used for describing Starobinsky inflation \cite{Ellis:2013xoa}. Further generalizations of the K\"ahler potential and the superpotential, keeping the no-scale structure and describing Starobinsky inflation in supergravity, are called Starobinsky-like in the literature. PBH production during inflation in the Starobinsky (modified) supergravity was investigated  in Refs.~\cite{Aldabergenov:2020bpt,Ishikawa:2021xya}. PBH production in the (modified) Cecotti-Kallosh model  \cite{Cecotti:2014ipa} of inflation in Starobinsky-like supergravity was initiated in Ref.~\cite{Aoki:2022bvj}.

The Starobinsky-like Einstein supergravity studied in Ref.~\cite{Aoki:2022bvj} has the following "no-scale" K\"ahler potential and the superpotential:
\be K=-3\ln \left(T+\bar{T}-|C|^2+\zeta\fracmm{|C|^4}{T+\bar{T}}\right)~, \quad  W=MC(T-1)+g(T),\label{W}
\ee
which include the new real parameter $\zeta$ and arbitrary holomorphic function $g(T)$.  The model of Ref.~\cite{Cecotti:2014ipa} arises in the case of $g(T)=0$. The case of $\zeta=g(T)=0$ defines the "no-scale" structure. However, it does not describe viable inflation because of an instability in the space of scalars (too small  $N$) so that the $\zeta$-dependent term is required for Starobinsky inflation near $C=0$ with the effective inflaton mass $M$. Therefore, it is possible to set $C=0$ in equations of motion, which greatly simplifies calculations of inflationary dynamics and PBH production.

The main idea of Ref.~\cite{Aoki:2022bvj} was to find an analytic function $g(T)$ in the superpotential, in order to describe PBH production during Starobinsky inflation by demanding the existence of the Ultra-Slow-Roll (USR) phase between two Slow-Roll (SR) phases of inflation, which leads to a large peak in the power spectrum of density perturbations that later collapse to PBH. In our models (\ref{W}), the USR phase arises via the so-called "iso-curvature pumping mechanism"  due to tachyonic instabilities in multi-field inflation, producing large iso-curvature perturbations that act as the source of large density perturbations that later gravitationally collapse into PBH, see Refs.~\cite{Gundhi:2018wyz,Gundhi:2020kzm} for details. 

It is worth mentioning that the existence of an USR phase in the supergravity model (\ref{W}) is by no means guaranteed. First, one has to suppress extra scalars in order to get single-field SR inflation (this problem was solved in Ref.~\cite{Aoki:2022bvj}). Second, one has to get good agreement (within $1\s$) with the Planck data in Eq.~(\ref{tilts}). Third, one has to produce a large peak in the power spectrum, which would generate PBH with masses beyond the Hawking evaporation limit of $10^{15}$ g, so that those PBH would survive in the current universe and may form dark matter. Third, as was demonstrated in Ref.~\cite{Aoki:2022bvj}, a linear combination of linear and exponential functions $g$ of $T$ does not do the job for any choice of the parameters. It was the reason why a quadratic polynomial $g(T)$ was considered in Ref.~\cite{Aoki:2022bvj} where it was found that fine-tuning of the parameters in a quadratic superpotential does lead to viable inflation and PBH production beyond the Hawking limit, though in agreement with the Planck-measured $n_s$ within $3\s$ but outside $1\s$. It follows from this fact that the model proposed and studied in Ref.~\cite{Aoki:2022bvj} may be easily falsified by more precise measurements of the $n_s$ tilt in the near future.

Required calculations and scanning of the parameter space in the models (\ref{W}) can only be performed numerically by using Mathematica and a dedicated code on the case-by-case basis.  We used the transport method and the open code provided in Ref.~\cite{Dias:2015rca}. 

It is more practical (for numerical calculations) to make Cayley transformation from the complex variable $T$ defined in the upper half-plane to another complex variable $Z$ defined on Poincar\'e disk, and redefine $C$ in favor of $S$ as follows:
\be \lb{cayley} T=\fracmm{1+Z}{1-Z}, \quad C= \fracmm{\sqrt{2}S}{1-Z}~~.
\ee
The inverse transformation reads
\be \lb{icayley}
Z=\fracmm{T-1}{1+T}, \quad S= \fracmm{\sqrt{2}C}{1+T}~~.
\ee

The scalar part of the Lagrangian in the model (\ref{W}) with $C=S=0$ is given by
\be \lb{kin}
\mathcal{L}=- \fracmm{3}{\left(1-|Z|^{2}\right)^{2}} \partial_{\mu} Z \partial^{\mu} \bar{Z}-V
\ee
with the scalar potential 
\begin{align} \lb{pot}
   V  =&\ \frac{M^2}{3} \fracmm{|Z|^2\left|1-Z\right|^{2}}{\left(1-|Z|^{2}\right)^{2}}+\frac{1}{24} \fracmm{|1-Z|^{6}}{1-|Z|^{2}}\left|\fracmm{dg}{dZ}\right|^{2}-\frac{1}{8} \fracmm{|1-Z|^{4}}{\left(1-|Z|^{2}\right)^{2}}\left[(1-Z)^{2} \fracmm{dg}{dZ} \bar{g}+(1-\bar{Z})^{2} \fracmm{d \bar{g}}{d \bar{Z}} g\right]
\end{align}
in terms of complex inflaton $Z$ and its complex conjugate $\bar{Z}$, with a holomorphic  function $g(Z)$ and its conjugate $\bar{g}(\bar{Z})$. The
canonical inflaton $\varphi$ appears in the parametrization 
\be \lb{parz} Z=re^{i\theta}={\rm{tanh}}  \fracmm{\varphi}{\sqrt{6}}\,e^{i\theta}
\ee
because the kinetic term in this parametrization takes the form
\be \lb{kinp}
\mathcal{L}=-\frac{1}{2} (\partial_{\mu}\varphi)^2-\frac{3}{4}{\rm{sinh}}^2 \fracmm{2\varphi}{\sqrt{6}}(\partial_{\mu}\theta)^2  
\ee
with sinflaton $\theta$. The two-dimensional target (field) space of this non-linear sigma-model has hyperbolic geometry with a constant negative curvature equal to $-4/3$. 

As was demonstrated in Ref.~\cite{Aoki:2022bvj} in the case of $g=0$, sinflaton $\theta$ is stabilized during SR inflation because of its heavy mass beyond the Hubble value, so that sinflaton dynamics and related iso-curvature perturbations can be ignored during SR inflation. However, it does not apply to the USR phase needed for PBH production, because of a tachyonic instability in the sinflaton direction for special choices of $g(Z)$ function, see next Sections. Given $\theta=g(Z)=0$ the scalar potential (\ref{pot}) in the parametrization (\ref{parz}) reduces to the Starobinsky potential (\ref{starp}), see Section 2 of Ref.~\cite{Aoki:2022bvj} for a detailed derivation.

When the function $g\neq 0$, both scalars should be taken into account.  The equations of motion in the flat FLRW universe are
\begin{align}
  &0=\ddot{\varphi}+3H\dot{\varphi}-\sqrt{\frac{3}{8}}\sinh\left(\sqrt{\frac{8}{3}}\varphi\right)\dot{\theta}^2+\partial_\varphi V~~,\nonumber\\
  &0=\ddot{\theta}+3H\dot{\theta}+2\sqrt{\frac{2}{3}}{\mathrm {coth}}{\left(\sqrt{\frac{2}{3}}\varphi\right)}\dot{\varphi}\dot{\theta}+\frac{2}{3}\mathrm{csch}^2\left(\sqrt{\frac{2}{3}}\varphi\right)\partial_\theta V~~,\label{eq2}\\
  &0=\frac{1}{2}\dot{\varphi}^2+\frac{3}{4}\sinh^2\left(\sqrt{\frac{2}{3}}\varphi\right)\dot{\theta}^2-3H^2+V~~,\nonumber
\end{align}
with the potential 
\begin{align} \lb{pot2}
  V=&\fracmm{M^2}{3}\fracmm{r^2(1-2r\cos\theta+r^2)}{(1-r^2)^2}+\fracmm{1}{24}\fracmm{(1-2r\cos\theta+r^2)^3}{1-r^2}\left|\fracmm{\mathrm{d}g}{\mathrm{d}Z}\right|^2\nonumber\\
    &-\fracmm{1}{8}\fracmm{(1-2r\cos\theta+r^2)^2}{(1-r^2)^2}\left[(1-re^{i\theta})^2\fracmm{\mathrm{d}g}{\mathrm{d}Z}\bar{g}
    +(1-re^{-i\theta})^2\fracmm{\mathrm{d}\bar{g}}{\mathrm{d}\bar{Z}}g\right]~~.
\end{align}

It is straightforward to derive the equations for perturbations by varying the equations of motion with respect to $\varphi$ and $\theta$ with the potential (\ref{pot}). However, those equations are long and not very illuminating, so that we do not quote them here.

\section{New ansatz for $g$-function}

In this paper we investigate the Starobinsky-like supergravity models (\ref{W}) with more general $g$-functions having the form
\begin{align}\lb{gfunc}
  g(Z)=M(g_0+g_1Z+g_2Z^2+g_3Z^3)
\end{align}
and four parameters $(g_0,g_1,g_2,g_3)$. Equation (\ref{gfunc}) is the most general renormalizable (Wess-Zumino) superpotential. Unlike Ref.~\cite{Aoki:2022bvj}, we assume $g_3\neq 0$. Actually, the latter condition leads to a rather complicated scalar potential from Eqs.~(\ref{pot}) and (\ref{pot2}).

The parameter $g_0$ is fixed in terms of the other parameters by demanding a Minkowski vacuum in the potential, which amounts to solving the
equations
\begin{align}
  \partial _\varphi V=\partial _\theta V=V=0~. \lb{g0}
\end{align}

Next, we impose the desired inflationary dynamics by demanding the effective single-field Starobinsky inflation during slow-roll, a saddle point in the potential below the first inflationary plateau and a sharp turn of the inflationary trajectory after the saddle point toward another plateau, in order to get three-phase (SR-USR-SR) inflation and PBH production during the USR phase. All that leads to restrictions on the free parameters $(g_1,g_2,g_3)$. We illustrate those features by a sample solution to the equations of motion (\ref{eq2}), which is shown in Fig.~\ref{qua_old}. The realized scenario is as follows: (i) during the first SR phase of inflation driven by scalaron $\varphi$, sinflaton $\theta$ is suppressed, (ii) when approaching the saddle point, the effective mass squared of sinflaton becomes negative, which signals a tachyonic instability leading to large iso-curvature perturbations fueling large density perturbations during the USR phase, (iii) after the USR phase, the second SR phase of single-field inflation begins with the sinflaton as the driver having the effective positive mass squared. The double SR inflation implies the existence of two plateaus in the Hubble function $H(t)$.

\begin{figure}[h]
  \begin{tabular}{ccc}
    \begin{minipage}[t]{0.45\hsize}
      \centering
      \includegraphics[keepaspectratio, scale=0.8]{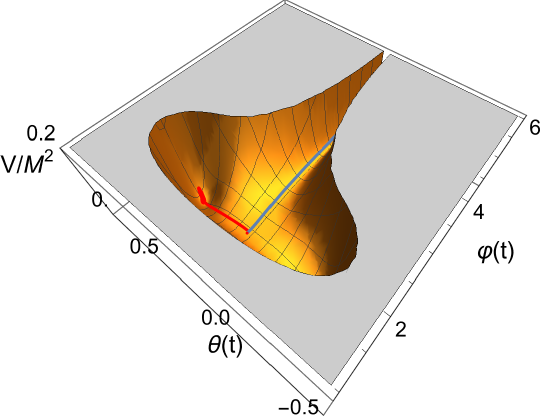}
      \subcaption{}
    \end{minipage} &
    \begin{minipage}[t]{0.45\hsize}
      \centering
      \includegraphics[keepaspectratio, scale=0.8]{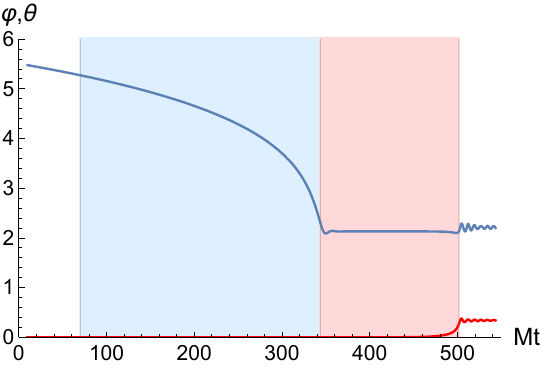}
      \subcaption{}
    \end{minipage} 
  \end{tabular}
  \caption{The shape of the sample potential with the inflationary trajectory (a), and the corresponding solution to the equations of motion with the  parameters $(g_0,g_1,g_2,g_3)\approx (14.3,-21,0,16)$ and the initial conditions $\varphi_0=5.5,~\theta_0=0$ with
  $\theta'=1.4\times 10^{-6}$ (b). The blue part of the  trajectory corresponds to the first SR stage of the inflation and the red part corresponds to the second SR stage of inflation in (a). The USR phase between  the SR phases is short (a few e-folds), being almost invisible on these plots. The blue line describes $\varphi$-evolution and the red line describes $\theta$-evolution in (b). }
  \lb{qua_old}
\end{figure}

The SR multi-field inflation parameters $\epsilon$ and $\eta_{\Sigma\Sigma}$ are defined by \cite{Dias:2015rca,Geller:2022nkr}
\begin{align}
  &\epsilon=-\fracmm{\dot{H}}{H^2}~,\quad \eta_{\Sigma\Sigma}=\abs{\fracmm{\mathcal{M}_B^A\Sigma_A\Sigma^B}{V}}~,\quad 
  \phi^A=\{\varphi, \theta\}~, \quad A,B,\ldots=1,2~,  \\
  &\Sigma^A=\fracmm{\dot{\phi}^A}{|\dot{\phi}|}~,\quad \mathcal{M}_B^A=G^{AC}\nabla_B\partial_CV-\mathrm{R}^A_{~CDB}\dot{\phi}^C\dot{\phi}^D~,\quad \abs{\dot\phi}^2=G_{AB}{\dot\phi}^A{\dot\phi}^B~,
\end{align}
in terms of the metric $G_{AB}$, the Christoffel symbols $\G^A_{BC}$, the related covariant derivatives $\nabla_A$ and  the 
Riemann-Christoffel curvature  $\mathrm{R}^A_{~CDB}$ in the field space associated with the non-linear sigma-model (\ref{kinp}),
where  $\partial_A$ are the usual derivatives with respect to $\phi^A$. The iso-curvature parameter is defined by \cite{Dias:2015rca,Geller:2022nkr}
\begin{align}
  \eta_{\Omega\Omega}=\fracmm{\mathcal{M}_B^A\Omega_A\Omega^B}{V}~, \quad
  \Omega^A=\fracmm{{\omega}^A}{|{\omega}|}~, \quad
  \omega =\dot{\Sigma^A}+\Gamma^A_{BC}\Sigma^B\dot{\phi}^C~~,\quad \abs{\omega}^2=G_{AB}{\omega}^A{\omega}^B~.
\end{align}

The Hubble function during SR inflation is determined by the potential via Friedmann equation. The CMB tilts (the scalar spectral index $n_s$ and the tensor-to-scalar ratio $r$) are simply related to the SR parameters in the first order with respect to perturbations as
\cite{Dias:2015rca,Geller:2022nkr}
\begin{align}\lb{titlsSR}
  n_s=1-6\epsilon+2\eta_{\Sigma\Sigma} \quad {\rm and} \quad r\leq 16\epsilon~,
\end{align}
which are to be evaluated at the CMB pivot scale $k_*=0.05~\mathrm{Mpc}^{-1}$.

The USR phase appears when $\e$ becomes very small$, \e\to +0$, which leads to a large enhancement (peak)  in the power spectrum of scalar perturbations. The SR conditions are violated in the USR phase with  $\eta_{\Sigma\Sigma}\equiv \eta >1$.

The start of the second SR stage of inflation is determined by the time when the parameter $\eta$ goes back to one (or when the parameter 
$\epsilon$ first reaches its maximum), while the end of the second stage is determined by the time when the parameter $\epsilon$ reaches one first. 
We set the total duration of inflation to $70$ e-folds. The target value of the second SR stage of inflation is chosen to be about 20 e-folds.

We used Mathematica for numerically computing the scalar potential and getting solutions to the equations of motion by randomly taking initial conditions for the scalars and scanning the parameter space $(g_1,g_2,g_3)$. The standard transport method and the Mathematica package (code) of Ref.~\cite{Dias:2015rca} were used to numerically calculate the  power spectrum of perturbations in our models. We refer to Ref.~\cite{Dias:2015rca} for details.

A dependence of our solutions upon initial conditions on $\varphi$ was found to be weak, which was not surprising because the Starobinsky inflationary solution is an attractor. Contrary to that, a dependence of our solutions upon initial conditions on sinflaton $\theta$ was found to be relevant because initial conditions $\theta_0$ and ${\theta_0}'$ determine the classical velocity $\theta'$ at the saddle point, while the $\theta'$ appears to be the key parameter for the power spectrum enhancement.~\footnote{The primes denote the time derivatives here.} 

When approaching the saddle point, the velocity $\theta'$ becomes very small so that quantum corrections become important. In principle, since  the $\theta'$ is not an independent parameter, its value at the saddle point may be calculated in quantum theory by averaging over quantum fluctuations in the stochastic approach, known as quantum diffusion~\cite{Pattison:2021oen,Tada:2023fvd,Vennin:2024yzl}. However, this task requires another framework and non-perturbative methods different from ours. Instead, we decided to scan possible values of $\theta'$ randomly  and determine whether any of them can lead to efficient PBH production. We found such values listed in Table 1, see the next Section. Those values may be falsified by quantum calculations.

In addition, we had to avoid approaching $\theta'=0$ in our calculations because of failure of numerical calculations of the power spectrum. Presumably, it was caused by the presence of resonances on sub-horizon scales, when a massive scalar oscillates at the bottom of the potential \cite{Pahud:2008ae,Chluba:2015bqa,Fumagalli:2021cel}. Because of that complication, the power spectrum was computed separately for each stage of inflation by using independent initial conditions, see Section 6 of Ref.~\cite{Aoki:2022bvj} for more details.
 
\section{Results} 

The most time-consuming part of our investigation was scanning the parameter space $(g_1,g_2,g_3)$ in a search for proper potential and deriving
the power spectrum and CMB observables $n_s$ and $r$ for some small values of $\theta'$ at the saddle point. As a result, we found 
the parameters $g_1$ and $g_3$  have to be non-vanishing and fine-tuned, whereas the parameter $g_2$ should  essentially vanish.  Actually, the parameters $g_1$ and $g_3$  were also related to each other due to the required shape of the potential, so that in practice we had to increase both $g_1$ and $g_3$ and cancel the negative effect of $g_2$ by a positive impact of $g_1$ during animation of the potential by varying the parameters.
Our best findings are collected in Table 1.

\begin{table}[ht]
  \centering
  \begin{tabular}{c c c c c c c c c c}
  \toprule
  $g_1$ & $g_2$ & $g_3$ & $g_0$ & $n_s$ & $r$ & $\theta '$ & $\Delta N_2$ & $M_{PBH}~({\rm g})$\\
  \midrule
  $-20.0$ & $0$ & $15.0$ & $13.9$ & $0.96081$ & $0.004085$ & $4.5\times10^{-5}$ & $19.7$ & $1.3\times10^{17}$  \\
  $-20.4$ & $0$ & $15.4$ & $14.1$ & $0.96079$ & $0.004078$ & $1.1\times10^{-5}$ & $19.9$ & $2.1\times10^{17}$  \\
  $-20.0$ & $0.2$ & $15.0$ & $13.8$ & $0.96088$ & $0.004067$ & $1.1\times10^{-5}$ & $19.7$ & $1.4\times10^{17}$  \\
  $-20.6$ & $0$ & $15.6$ & $14.2$ & $0.96081$ & $0.004072$ & $5.5\times10^{-6}$ & $19.9$ & $2.2\times10^{17}$  \\
  $-20.8$ & $0$ & $15.8$ & $14.2$ & $0.96092$ & $0.004046$ & $2.9\times10^{-6}$ & $19.8$ & $1.8\times10^{17}$  \\
  $-21.0$ & $0$ & $16.0$ & $14.3$ & $0.96089$ & $0.004046$ & $1.4\times10^{-6}$ & $19.9$ & $2.2\times10^{17}$  \\
  \bottomrule
  \end{tabular}
  \captionsetup{width=.9\linewidth}
  \caption{The values of the parameters $(g_1,g_2,g_3,g_0)$ and the corresponding CMB observables $(n_s,r)$, with a small kick 
  $\theta'$ at the saddle point, the duration $\D N_2$ of the second SR phase of inflation, and the resulting PBH masses. The initial conditions are   chosen to be $\varphi_0=5.5$ and $\theta_0=0$ with the vanishing velocities. The pivot scale $k_*=0.05~\mathrm{Mpc}^{-1}$ corresponds to   $63.79$ e-folds, while the total duration of inflation is 70 e-folds.}
  \lb{cubtab1}
\end{table}

The shape of the scalar potential and the inflationary trajectory with the fine-tuned parameters  $(g_1,g_2,g_3)=(-21,0,16)$ are the same as in Fig.~\ref{qua_old}. The value of the potential height in the saddle point is $V/M^2\approx 0.05$. The Hubble function with two plateaus corresponding to two SR phases of inflation is given in Fig.~\ref{cubhub}. The inflaton mass $M$ is approximately $0.8\times10^{-5}$, and the CMB scales are around $2.0\times 10^{-9}$.

\begin{figure}[t]
  \centering
  \begin{minipage}[t]{0.45\hsize}
    \includegraphics[keepaspectratio, scale=1.0]{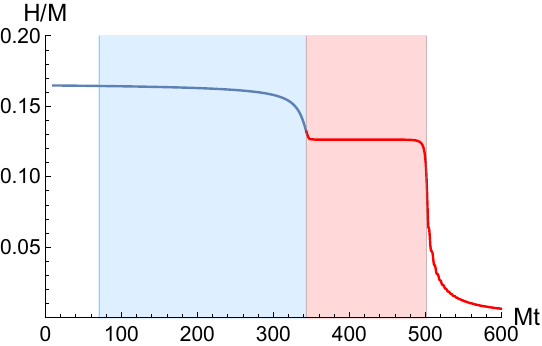}
  \end{minipage}
  \caption{The plot of the Hubble function for the parameters $g_1=-21$, $g_2=0$ and $g_3=16$. The blue shaded region shows the first SR stage of  inflation and the red shaded region shows the second SR stage of inflation.}
  \lb{cubhub}
\end{figure}

The SR parameters $\e$ and $\eta$ are displayed in Fig.~\ref{cubslow} where the plot (a) shows the presence of the USR phase because of
very small values of $\e$ of the order $10^{-8}$, whereas the plot (b) demonstrates a violation of the SR conditions in the USR phase because of 
$\eta>1$. The values of $\eta_\mathrm{CMB}$ at CMB scales are around $0.015$ and increase to $0.2$ at the end of the first SR stage of inflation. 
The values of $\eta_\mathrm{USR}$ during the USR phase are between $3$ and $4$. The duration of the USR phase is $5.5$ 
e-folds.
\begin{figure}[t]
  \begin{tabular}{ccc}
    \begin{minipage}[t]{0.45\hsize}
      \centering
      \includegraphics[keepaspectratio, scale=0.8]{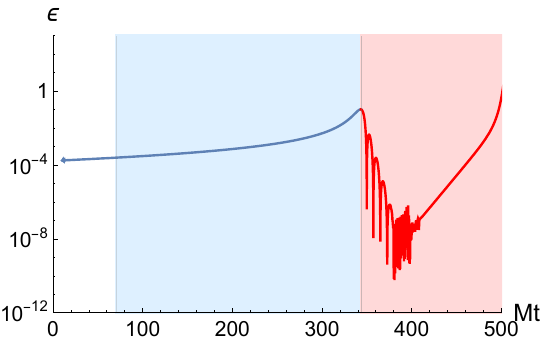}
      \subcaption{}
    \end{minipage} &
    \begin{minipage}[t]{0.45\hsize}
      \centering
      \includegraphics[keepaspectratio, scale=0.8]{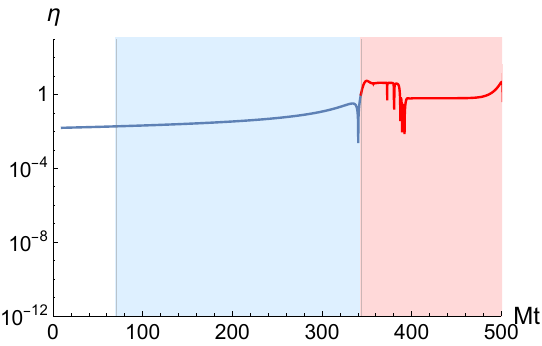}
      \subcaption{}
    \end{minipage} 
  \end{tabular}
  \caption{The plots of the SR parameters $\epsilon$ in (a) and $\eta$ in (b) for $g_1=-21$, $g_2=0$ and $g_3=16$. The blue shaded region shows the first SR stage of inflation and the red shaded region shows the second SR stage of inflation.}
  \lb{cubslow}
\end{figure}

The derived values of the CMB tilt $n_s$ in Table 1 are within the $1\sigma$ range of the Planck-observed value in Eq.~(\ref{cmbt}), while the derived
values of the tensor-to-scalar ratio $r$ in Table 1 are well below the current observational bound in Eq.~(\ref{cmbt}).

The power spectrum is given in Fig.~\ref{cubpowfit} for the fine-tuned parameters collected in Table 1. The peaks in the power spectrum have heights
between $10^{-2}$ and $10^{-3}$ which are about 7 to 6 orders of the magnitude higher than the CMB level of the order $10^{-9}$ (on the left of the plot), respectively.

\begin{figure}[h]
  \begin{minipage}[t]{0.45\hsize}
    \includegraphics[keepaspectratio, scale=1.0]{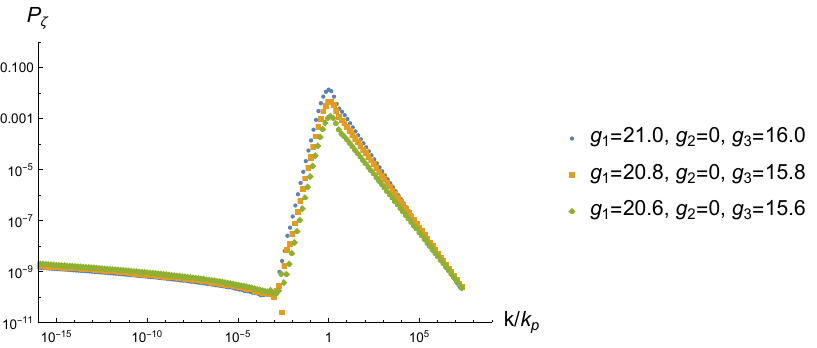}
  \end{minipage}
  \caption{The power spectrum of scalar perturbations for the parameter sets in Table.~\ref{cubtab1}. The normalization scale
  $k_p$ corresponds to the peak at $k_p\approx 10^{14}\,{\rm Mpc}^{-1}$.}
  \lb{cubpowfit}
\end{figure}
 
Given the power spectrum, the masses of PBH originating from large over-densities can be estimated as follows \cite{Pi:2017gih}:
\be \lb{cubmass}
  M_{\mathrm{PBH}}\simeq\fracmm{1}{H(t_*)}\mathrm{exp}\left[2(N_{\mathrm{end}}-N_*)+\int_{t_*}^{t_{\mathrm{exit}}}\epsilon(t)H(t)dt\right]~,
\ee
where $t_*$ is the time when the first SR stage of inflation ends and $t_{\mathrm{exit}}$ is the time when the CMB pivot scale 
$k_*=0.05~\mathrm{Mpc}^{-1}$ exits the horizon. The leading term in the exponential is given by $2\D N_2=2(N_{\mathrm{end}}-N_*)$, whereas the
second term in the exponential is sub-leading. The value of $\Delta N_2$ is close to $20$ in our model, which results in the PBH masses 
of the order $10^{17}$ g given in Table 1. 

The dependence of  $\Delta N_2$  upon $\theta'$ is shown in Fig.~\ref{cubini}.

\begin{figure}[h]
  \centering
  \begin{minipage}[t]{0.45\hsize}
    \includegraphics[keepaspectratio, scale=1.0]{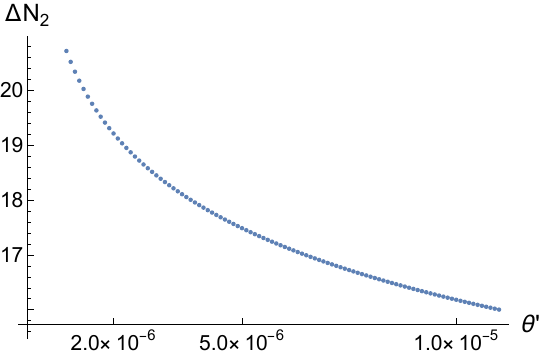}
  \end{minipage}
  \caption{The impact of quantum kick $\theta'$ on $\Delta N_2$ for $g_1=-21$, $g_2=0$ and $g_3=16$.}
  \lb{cubini}
\end{figure}

The power spectrum enhancement (large peak) describes large scalar perturbations near the critical point, which are essentially independent upon initial conditions but are sensitive to the kick velocity $\theta'$ at the saddle point.

We compared our result for the peak in the power spectrum of Fig.~\ref{cubpowfit} to the log-normal fit of the peak \cite{Pi:2020otn,Domenech:2021ztg, Frolovsky:2023xid},
\be \lb{logn}
  \mathcal{P}_\zeta(k)=\fracmm{\mathcal{A}_\zeta}{\sqrt{2\pi}\Delta}\mathrm{exp}\left[-\fracmm{\mathrm{ln}^2(k/k_p)}{2\Delta^2}\right], 
\ee
which has only two adjustable parameters, the amplitude $\mathcal{A}_\zeta$ and the width $\D$ of the peak at the location scale $k_p$. A comparison of the power spectrum and the log-normal fit is shown in Fig.~\ref{cubpow}.

\begin{figure}[h]
  \centering
  \begin{minipage}[t]{0.45\hsize}
    \includegraphics[keepaspectratio, scale=1.0]{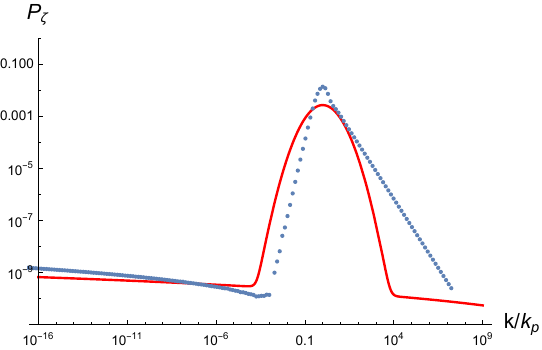}
  \end{minipage}
  \caption{The plot of the power spectrum (blue) in the case $(g_1,g_2,g_3)=(-21,0,16)$, against the log-normal fit (red) with the amplitude 
  $\mathcal{A}_{\zeta}=0.01$ and the width $\Delta$=1.5.}
  \lb{cubpow}
\end{figure}

We used the Press-Schechter formalism \cite{Press:1973iz} to estimate the PBH fraction of DM in our model from the power spectrum shown in Fig.~\ref{cubpowfit} with the parameters given in Table 1. The relevant equations can be found in Ref.~\cite{Inomata:2017okj}, see also Section 6 of Ref.~\cite{Aldabergenov:2022rfc}.~\footnote{The Press-Schechter formalism should be considered with a grain of salt because it was found to be unreliable~\cite{Franciolini:2018vbk,Cruces:2018cvq,Figueroa:2021zah}.}  We found that our results strongly depend upon the values of the model parameters and the density threshold $\delta_c$ for PBH formation. When using the fine-tuned
parameters in Table 1 and $1/3\leq \delta_c \leq 2/3$, we got the $f_{\rm PBH}=\Omega_{\rm PBH}/\Omega_{\rm DM}$  between $10^{-3}$ and $1$. The plot in Fig.~\ref{cubpow} was obtained with $\delta_c\approx 0.55$ leading to the peak height $P_{\z,{\rm max.}}\approx 0.016$. It is worth mentioning that even a small contribution of PBH to DM  can play an important cosmological role \cite{Carr:2020gox}.

\section{Gravitational waves induced by PBH production}

PBH production leads to induced (stochastic) gravitational waves (GW) that are different from primordial GW caused by inflation. The current energy density  fraction of the PBH induced GW can be computed in the second order with respect to perturbations, and the result reads  \cite{Espinosa:2018eve,Kohri:2018awv}
\begin{align}
  \Omega_{\mathrm{GW}}(k)=&\fracmm{c_g\Omega_{r,0}}{36}\int^{\fracmm{1}{\sqrt{3}}}_{0}  \,\mathrm{d}d \int^{\infty}_{\fracmm{1}{\sqrt{3}}}  \,\mathrm{d}s \left[\fracmm{(d^2-1/3)(s^2-1/3)}{s^2-d^2}\right]^2\times\nonumber\\
  &\mathcal{P}_\zeta\left(\fracmm{k\sqrt{3}}{2}(s+d)\right)\mathcal{P}_\zeta\left(\fracmm{k\sqrt{3}}{2}(s-d)\right)\left[
  \mathcal{I}_c(d,s)^2+\mathcal{I}_s(d,s)^2\right]~,\lb{eqp}
\end{align}
where the functions $\mathcal{I}_c(d,s)$ and $\mathcal{I}_s(d,s)$ are given by
\begin{align}
  \mathcal{I}_c(d,s)&=-36\pi\fracmm{(s^2+d^2-2)^2}{(s^2-d^2)^3}\theta(s-1)~,\nonumber\\
  \mathcal{I}_s(d,s)&=-36\fracmm{s^2+d^2-2}{(s^2-d^2)^2}\left[\fracmm{s^2+d^2-2}{(s^2-d^2)}\mathrm{ln}
  \abs{ \fracmm{d^2-1}{s^2-1}}+2\right]~,
\end{align}
with $\Omega_{r,0}\sim8.6\times10^{-5}$ being the current energy density fraction of radiation. The $\theta(s-1)$ is the step function and $c_g\approx 0.4$.

A numerical calculation of Eq.~(\ref{eqp}) with our power spectrum in Fig.~\ref{cubpow} yields the result given in Fig.~\ref{gwfit2} (in blue).
The numerical plot can be approximated by the lognormal fit (see Fig.~\ref{gwfit2}, in red)
\be \lb{logngw}
  \O_{\rm GW}(k)=\fracmm{A_{\rm GW}}{\sqrt{2\pi}\sigma}\mathrm{exp}\left[-\fracmm{\mathrm{ln}^2(k/k_p)}{2\sigma^2}\right]
\ee
with the amplitude $A_{\rm GW}\approx 1.2\cdot 10^{-8}$ and the width $\sigma\approx 1.06$, so that
$\Omega_{\mathrm{GW}}^{\rm peak}(k)\sim10^{-5}\mathcal{P}^2_\zeta(k)$ near the peak, in agreement with the estimates in Refs.~\cite{Pi:2020otn,Domenech:2021ztg}. 

\begin{figure}[h]
  \centering
  \begin{minipage}[t]{0.45\hsize}
    \includegraphics[keepaspectratio, scale=1.0]{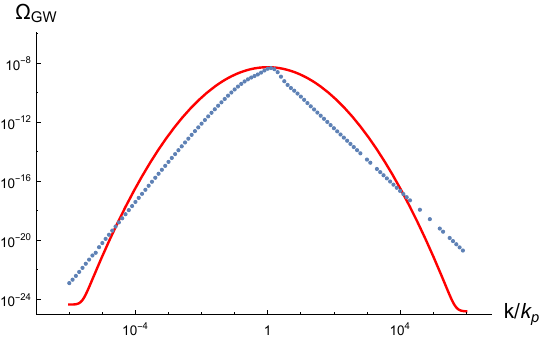}
  \end{minipage}
  \caption{The GW density fraction (blue) in the case $(g_1,g_2,g_3)=(-21,0,16)$ against the log-normal fit (red) of the peak in Eq.~(\ref{logngw}).}
  \lb{gwfit2}
\end{figure}

The induced GW frequencies are related to the PBH masses as \cite{DeLuca:2020agl}
\be \lb{gwf}
  f_p\approx5.7\left(\fracmm{M_\odot}{M_{\mathrm{PBH}}}\right)^{1/2}10^{-9}~\mathrm{Hz},
\ee
where $M_\odot\approx 2\cdot 10^{33}$ g is the mass of the Sun. In our models, the PBH masses are about $2\cdot 10^{17}g$, which results in the GW frequency $f_p\approx 0.6$~Hz that is much higher than the GW frequencies between 3 and 400 nHz 
detected by NANOGrav \cite{NANOGrav:2023gor}. A detection of a GW peak with a near Hz frequency may provide observational
support to Starobinsky-like supergravity, as was argued in Ref.~\cite{Basilakos:2023xof}, if such peak is due to the generation of secondary GW. 

A better comparison of our predictions with future GW observations is possible by plotting the induced  GW density in our model against the expected sensitivity curves in the space-based gravitational interferometers LISA \cite{LISA:2017pwj,Smith:2019wny}, TianQin \cite{TianQin:2015yph}, Taiji \cite{Gong:2014mca,Ruan:2018tsw} and DECIGO \cite{Kudoh:2005as}, see Fig.~\ref{sensi}, where we have used Refs.~\cite{Thrane:2013oya,Schmitz:2020syl,Aldabergenov:2020yok} for the colored curves. Therefore, the upcoming space-based
experiments are expected to be sensitive to stochastic GW in the frequencies between $10^{-3}$ and $10^{-1}$ Hz, while the upper left part of the black curve in Fig.~\ref{sensi} belongs to this frequency range.

\begin{figure}[h]
  \centering
  \begin{minipage}[t]{0.45\hsize}
    \includegraphics[keepaspectratio, scale=1.0]{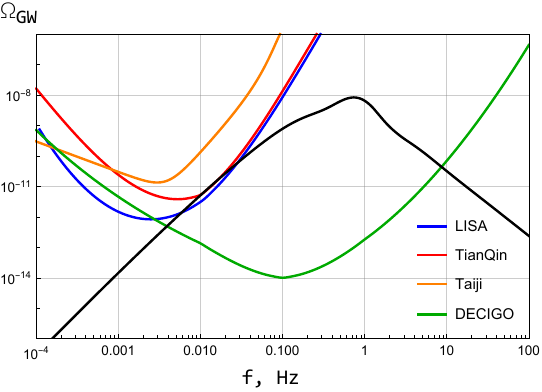}
  \end{minipage}
  \caption{The stochastic GW density induced by the power spectrum enhancement of scalar perturbations in our model (in black) against the expected sensitivity curves for the future space-based GW experiments (in color).}
\lb{sensi}
\end{figure}

\section{Conclusion}

The high scale of Starobinsky inflation naturally implies its local supersymmetrization in the context of four-dimensional supergravity,
while high-scale supersymmetry breaking avoids the Large Hadron Collider constraints on (low-scale) supersymmetry. It is well known that Starobinsky inflation can be realized in the standard (Einstein) supergravity  coupled to matter, see e.g., 
Ref.~\cite{Ellis:2013xoa}, while it is also possible in the modified Starobinsky supergravity \cite{Ketov:2023ykf} without significant fine-tuning.

More efforts are needed to realize efficient production of PBH with the masses beyond the Hawking evaporation limit during
inflation in good agreement with CMB observations because of common tension with the well-measured value of the CMB tilt $n_s$
of scalar perturbations. The necessity to produce a significant enhancement of the power spectrum of scalar perturbations during
inflation usually drives the predicted value of $n_s$ below its Planck-measured value, thus demanding fine-tuning of the parameters
in the potential. We do not think that fine-tuning itself is necessarily a problem because both inflation and (presumably) PBH production were unique (non-reproducible) events in the Universe.

The existence of viable inflation with efficient PBH production in the Starobinsky-like supergravity models is not guaranteed because of the constraints imposed by local supersymmetry. For instance, as was shown in Ref.~\cite{Aoki:2022bvj}, any linear combination of holomoprphic quadratic polynomials and exponentials of $Z$ in the superpotential  cannot solve the task within $1\sigma$ agreement  with the known value of $n_s$. Nevertheless, we demonstrated in this paper that it is possible by exploiting the cubic Wess-Zumino-type (renormalizable) superpotentials. The $g$-function used in Ref.~\cite{Aoki:2022bvj} 
was 
\be \lb{gqua}
 g(Z)=M(g_0+g_1Z+g_2Z^2)~.
\ee
The best choice of the coefficients $(g_0,g_1,g_2)\approx(1,-1,2)$ was shown to lead to the PBH masses beyond $10^{15}$ g within $3\sigma$  (but outside $1\sigma$) agreement with the $n_s$ value in Eq.~(\ref{cmbt}), whereas  demanding $1\sigma$ agreement with the CMB tilt $n_s$ was shown to drive the PBH masses below $10^{15}$ g  \cite{Aoki:2022bvj}.  In this paper, we found that the quadratic term should be absent, $g_2=0$, and the cubic term is essential for success, $g_3\neq 0$, see Table 1.

Our perturbative approach has its limitations, and we clearly indicated where they come from. Having identified the key parameter 
$\theta'$ relevant for efficient PBH production in our classical model, instead of trying to compute $\theta'$ in quantum theory we took its random values in our investigation. The required values of $\theta'$ were found to be small and, hence, may be falsified by averaging over
 quantum fluctuations in a non-perturbative approach. The significance of $\theta'$ is known in the literature, see e.g., Ref.~\cite{Clesse:2015wea}. The validity of the perturbation theory for  large density perturbations was discussed in Refs.~\cite{Kristiano:2022maq,Choudhury:2023rks,Firouzjahi:2023aum,Firouzjahi:2023ahg,Saburov:2024und}.

We briefly mentioned about our results for the PBH abundance in DM by using the "standard"  Press-Schechter method. However, as was shown in the literature~\cite{Franciolini:2018vbk,Cruces:2018cvq,Figueroa:2021zah}, this method is inadequate, so that we skipped details of our calculation. It was also pointed out in the literature \cite{Dvali:2020wft,Michel:2023ydf,Alexandre:2024nuo,Thoss:2024hsr} that the standard semiclassical result for black hole evaporation may be relaxed below $10^{15}$ g. Should this be the case, fine-tuning of the parameters for efficient PBH production in our model can be considerably relaxed.

Starobinsky inflation also provides the universal reheating mechanism \cite{Gorbunov:2010bn}. However, reheating after inflation and its possible impact on CMB are beyond the scope of this paper, see  however Refs.~\cite{Tsujikawa:1999iv,Fu:2019qqe,Ema:2024sit}.

The issue of ultra-violet (UV) completion of our supergravity models is beyond the scope of this investigation. To this end, we confine ourselves to a few comments.  Though very little is known for sure about quantum gravity, when assuming the quantum
gravity scale to be given by the Planck scale, the Starobinsky model and its supergravity extensions are on the safe side against quantum gravity corrections  because the scale of Starobinsky inflation is well below the Planck scale that is simultaneously the UV-cutoff scale for the Starobinsky gravity (\ref{stara}). When assuming the Swampland conjectures \cite{Palti:2019pca} that imply
a significantly lower quantum gravity scale, quantum gravity may have significant impact on our models so that their UV embedding is needed. It was recently argued in the literature \cite{Brinkmann:2023eph,Lust:2023zql} that it is difficult to realize Starobinsky inflation in superstring theory, both as regards Starobinsky gravity (\ref{stara}) and the Starobinsky potential (\ref{starp}),
as well as to reconcile the Starobinsky model   (\ref{stara})  with the Swampland program.

We think that inflation has happened, and we know that black holes exist, though we do not know whether some of them are primordial or not. In any case, inflation, supersymmetry and PBH genesis are the important windows into very high energy physics,
whose investigation put constraints on our understanding of the origin of the Universe.

\section*{Acknowledgements}

RI and SVK were supported by Tokyo Metropolitan University. SVK was also supported by the World Premier International Research Center Initiative (WPI Initiative), MEXT, Japan, the Japanese Society for Promotion of Science under the grant No.~22K03624, and the Development Program  Priority-2030-NIP/EB-004-375-2023 of Tomsk Polytechnic University.

The authors thank four referees for useful suggestions and critical remarks. One of the authors (SVK) is grateful to Ignatios Antoniadis, Shuntaro Aoki, Sayantan Choudhury, Guillem Domenech, David I. Kaiser, Jason Kristiano, Florian K\"uhnel, Kazunori Kohri, Edward W. Kolb, Theodoros Papanikolaou, Takahiro Terada, Burt Ovrut, Shi Pi, Misao Sasaki and Alexei Starobinsky for discussions and correspondence.

This paper is devoted to memory of late Alexei Starobinsky, one of the greatest cosmologists of all times.

\bibliography{Bibliography}{}
\bibliographystyle{utphys}

\end{document}